\documentclass[aps,prb,twocolumn,superscriptaddress,showpacs]{revtex4-1}
\usepackage{graphicx}
\usepackage{calc}
\usepackage{bm}

\begin{document}

\title{Local investigation of the energy gap within the incompressible strip  in the quantum Hall regime}

\author{E.V.~Deviatov}
\email[Corresponding author. E-mail:~]{dev@issp.ac.ru}
 \affiliation{Institute of Solid State
Physics RAS, Chernogolovka, Moscow District, 142432, Russia}

\author{A.~Lorke}
\affiliation{Laboratorium f\"ur Festk\"orperphysik, Universit\"at
Duisburg-Essen, Lotharstr. 1, D-47048, Duisburg, Germany}

\author{G.~Biasiol}
\affiliation{Laboratorio TASC, CNR-IOM, Area Science Park, I-34149 Trieste, Italy}

\author{L.~Sorba}
\affiliation{NEST, Istituto Nanoscienze and Scuola Normale Superiore, I-56127 Pisa, Italy}

\author{W.~Wegscheider}
\affiliation{Laboratory for Solid State Physics, ETH Z\"urich, CH-8093 Z\"urich, Switzerland}

\date{\today}

\begin{abstract}
We experimentally study the energy gap within the incompressible strip at local filling factor $\nu_c=1$ at the quantum Hall edge for samples of very different mobilities. The obtained results indicate strong enhancement of the energy gap in comparison to the single-particle Zeeman splitting. We identify the measured gap as a mobility gap, so a pronounced experimental in-plane magnetic  field dependence can both be attributed to the spin effects as well as to the change in the energy levels broadening.
\end{abstract}


\maketitle

The unique properties of a two-dimensional electron gas (2DEG) in the quantum Hall (QH)   regime originate from the presence of the energy gap at the Fermi level~\cite{prange}. The situation is especially interesting if only one spin-split sublevel of the lowest Landau level is occupied, i.e. at bulk filling factor $\nu=1$. In high magnetic fields, the ground state of the system is fully spin-polarized, while the elementary excitations are characterized by a spin flip~\cite{ando}. The exchange energy of a single spin-flip excitation is supposed to be responsible for a serious enhancement of the excitation energy in comparison to the single-particle Zeeman splitting~\cite{ando,nichols,aristov,kulik}.  This enhancement is  sensitive~\cite{ando} to the excitation wave vector $k$. It is widely accepted, that in   transport experiments~\cite{nichols,aristov,shmeller,khrapai} $k=\infty$ excitations are tested, while optical investigations~\cite{kulik} allow to probe excitations at different $k$. In  low magnetic fields, the possibility of complicated spin textures (skyrmions) formation is still     debated~\cite{shmeller,khrapai}.

The situation is even more intriguing at the sample edge. Any real edge profile is smooth, so the electron liquid represents a structure of   compressible and incompressible strips~\cite{shklovskii}. Every incompressible strip can be characterized by a constant local filling factor $\nu_c<\nu$. In  analogy with the bulk, the electron liquid is spin polarized within the $\nu_c=1$ incompressible strip, which should produce a local enhancement of the energy gap. It is not known {\em ab initio} how to extend the bulk results to the $\nu_c=1$ incompressible strip, because of, e.g.,  higher disorder at the edge, potential drop within the incompressible strip, etc.  Experimental verification is still missing, because the traditional methods (activation~\cite{nichols,shmeller},  magnetocapacitance~\cite{aristov,khrapai}, and optics~\cite{kulik}) are not suitable for   local investigations.

Here, we experimentally study the energy gap within the incompressible strip at local filling factor $\nu_c=1$ at the quantum Hall   edge for samples of very different mobilities. The obtained results indicate strong enhancement of the energy gap in comparison to the single-particle Zeeman splitting. We identify the measured gap as a mobility gap, so a pronounced experimental in-plane magnetic  field dependence can both be attributed to the spin effects as well as to the change in the energy levels broadening.

The samples are fabricated from  three GaAs/AlGaAs heterostructures with different carrier concentrations and mobilities, grown by molecular beam
epitaxy (MBE) in three different MBE machines. Two of them  (A and B) contain a 2DEG located 200~nm below the surface. These wafers are similar in the carrier density (about 1.6 $\cdot 10^{11}$cm$^{-2}$), but strongly differ in the 2DEG mobilities. The 2DEG mobility at 4K is  $5.5 \cdot 10^{6}  $cm$^{2}$/Vs for the wafer A, while it is equal to 1.93 $\cdot 10^{6}$cm$^{2}$/Vs  for the wafer B. For the lower   quality heterostructure C the  relevant parameters are 70~nm, 800 000 cm$^{2}$/Vs and 3.7 $\cdot 10^{11}  $cm$^{-2}$.

\begin{figure}
\includegraphics*[width=0.85\columnwidth]{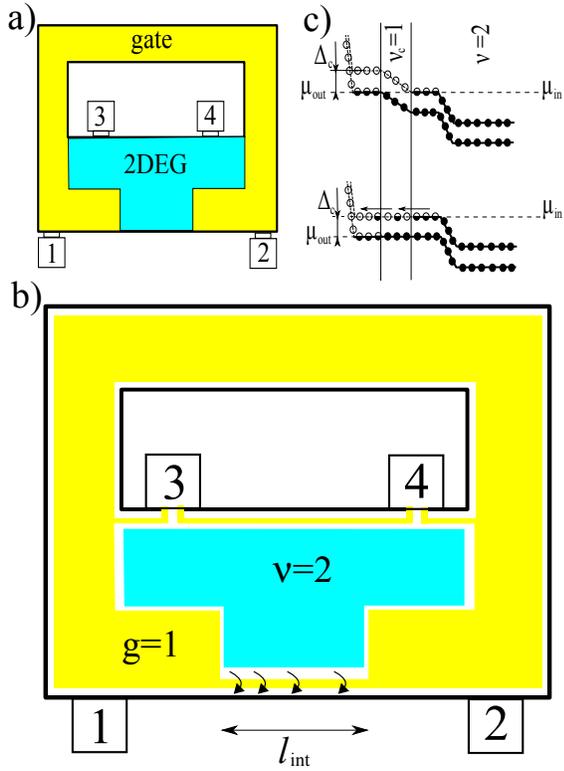}%
\caption{ (Color online) (a) Schematic diagram of the sample (not in the scale). The etched mesa edges are shown by thick solid lines. Light yellow (light gray) areas indicate the split-gate, which forms narrow ( 5-10~$\mu$m width) gate-gap region at the outer mesa edge. Light green (gray) area indicates uncovered 2DEG.  Ohmic contacts are denoted by bars with numbers. (b) Structure of the compressible (white) and incompressible (color) regions of the electron liquid in the sample. The bulk is at the filling factor $\nu=2$ (uncovered) or at the filling factor $g=1$ (under the gate). The compressible/incompressible strip structure is shown at the sample edges. Arrows indicate electron transport across $\nu_c=1$ incompressible strip in the gate-gap region. (c) Schematic diagram of  the  energy levels in the gate-gap region. Filled circles represent the fully occupied electron states in the incompressible strip  and in the bulk. Half-filled circles indicate the partially occupied electron states in the compressible strips. Pinning of the energy levels to the Fermi level (shot-dash) is shown in the compressible regions at electrochemical potentials $\mu_{out}$ and $\mu_{in}$.  Open circles are for the empty states. $\Delta_c$ is the potential jump in the $\nu_c$ incompressible strip at the equilibrium ($\mu_{out}=\mu_{in}$, no electrochemical imbalance is applied across the incompressible strip). Flat-band condition appears at the electrochemical potential imbalance $\mu_{in}-\mu_{out}=\Delta_c$. Arrows indicate a new way for  electrons  along the energy level.
\label{sample}}
\end{figure}

The samples are patterned in the quasi-Corbino sample
geometry~\cite{alida}, see Fig.~\ref{sample} (a). Each sample has a macroscopic ($\sim 0.5\times 0.5\mbox{mm}^2$) etched region inside, providing topologically independent inner and outer  mesa edges (the so called Corbino topology). Ohmic contacts are placed at the mesa edges.   A split-gate, encircling the etched region, is used to connect inner and outer mesa edges in a controllable way. It forms a narrow gate-gap region at the outer mesa edge. The gate-gap width equals to 5~$\mu$m  for samples A1, B, C, while it is twice wider (10~$\mu$m) for the sample A2.

In a quantizing magnetic field, at bulk filling factor $\nu=2$, we deplete the 2DEG under the gate to a lower filling factor $g=1$. A structure of compressible and incompressible strips emerges~\cite{shklovskii} at both mesa edges and along the gate edge, see Fig.~\ref{sample} (b). A filling factor $g$ under the gate is chosen to coinside with the local filling factor $\nu_c=1$ within the incompressible strip in the gate-gap region at the outer mesa edge. In this case, two compressible strips in the gate-gap originate from different (inner and outer) Ohmic contacts, being at their electrochemical potentials.  Applying dc bias between Ohmic contacts at outer and inner edges leads to an electrochemical potential imbalance across the incompressible strip at local filling factor $\nu_c=g=1$ in the gate-gap, see Fig.~\ref{sample} (b) and (c).

In the present experiment,  dc bias is applied to the outer contact with respect to the inner one. Schematic diagram of   energy levels in the gate-gap region is depicted in Fig.~\ref{sample} (c). Because of negative electron charge, the positive bias decreases the potential jump at $\nu_c=1$, allowing the flat-band situation at $eV=eV_{th}=\Delta_c$. If the gate-gap width $l_{int}$ is smaller than the characteristic equilibration length~\cite{mueller} $l_{eq}>100 \mu$m, the flat-band regime should reveal itself by a strong rising of the current, since no potential barrier exists between the compressible strips in the gate-gap. At higher biases, the excessive imbalance will be easily equilibrated until reaching the flat-band regime, which results in  the equilibrium slope of the positive $I-V$ branch above $eV=eV_{th}$. As a result, the positive branch of the $I-V$ curve allows the local determination of the energy gap within the $\nu_c$ incompressible strip at the sample edge~\cite{alida}.

The experimental data, presented here, are obtained from samples of different quality (A1, B, C) and with different gate-gap widths (A1 and A2). They are independent from the cooling cycle. All measurements are performed in a dilution refrigerator with base temperature 30~mK, equipped with a superconducting solenoid.       Standard magnetocapacitance and magnetotransport measurements were performed to characterize the electron system under the gate and in the ungated region.

\begin{figure}
\includegraphics[width= \columnwidth]{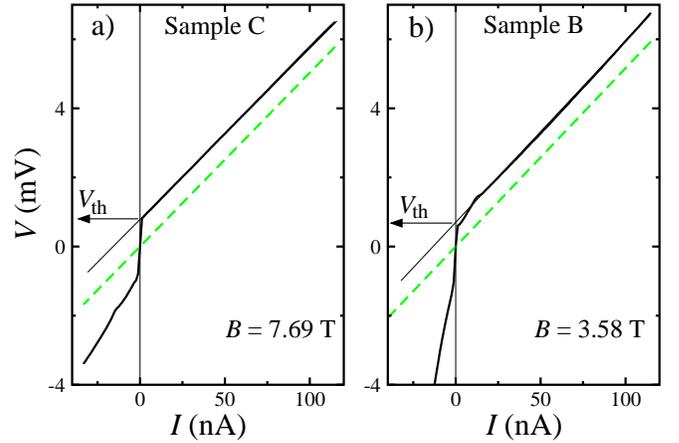}%
\caption{Examples of $I-V$ curves (solid) for the filling factor combination $\nu=2, g=1$ for samples with different electron concentrations and mobilities ((a) and (b) are for the samples C and B respectively).  The positive branch of the experimental curve is linear above the threshold ($V>V_{th}$, see dash for comparison). The threshold voltage $V_{th}$ is determined by linear extrapolation to zero current (thin lines). Normal magnetic fields equal to 3.58~T (b) and 7.69~T (a). \label{IV21}}
\end{figure}

Typical $I-V$ curves for transport across the integer incompressible strip $\nu_c=1$  are presented in  Fig.~\ref{IV21} for samples with different electron concentrations and mobilities. The experimental $I-V$ curve is strongly non-linear and asymmetric. The positive $I-V$ branch changes its slope at the finite threshold voltage $V_{th}$, the branch is linear above $V_{th}$ in a wide voltage range, see Fig.~\ref{IV21}. The negative branch  is strongly non-linear.  The threshold voltage $V_{th}$ can be obtained with high accuracy from the extrapolation of the positive linear branch to zero.

\begin{figure}
\includegraphics[width=\columnwidth]{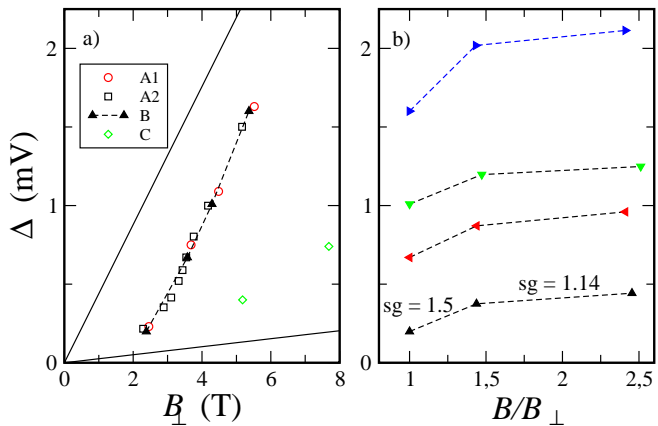}%
\caption{Energy gap in the $\nu_c=1$ incompressible strip as determined from the threshold voltage $V_{th}$ of experimental $I-V$ curves. (a) $\Delta_c$ is presented as a function of normal magnetic field for different samples (symbols). The dashed line is a guide to the eye. Solid lines demonstrate the bulk exchange-enhanced gap at $\nu=1$ ($g^*=7.3$ as it is measured in Ref.~\protect~\onlinecite{nichols}), and  a single-particle Zeeman splitting in GaAs. (b) In-plane field dependence of   $\Delta_c$ for the sample B in different fixed normal fields: $B_\perp=2.38$~T (up triangles), $B=3.58$~T (left triangles), $B=4.29$~T (down triangles), $B=5.36$~T (right triangles). Dash demonstrates different slopes $sg$ in low and high in-plane fields (see the text).    \label{gap}}
\end{figure}

The $\nu_c=1$ energy gap values, obtained as a threshold voltage for the positive $I-V$ branch are presented in Fig.~\ref{gap}. Accuracy of the gap determination procedure can be evaluated as 0.05~meV, which is roughly the size of the symbol in Fig.~\ref{gap}. Our experimental technique applies some restrictions on the normal magnetic field values: the field should correspond to the QH state in the bulk with the bulk filling factor $\nu>1$. On the other hand, the magnetic field should be high enough to fulfill $l_{int}<<l_{eq}$ condition. Within these restrictions,  the experimental data cover a significant field range,  allowing the following conclusions:

(i) The data obtained from  high-quality wafers A and B form a single dependence,  despite the mobilities differ by a factor 2.   In contrast, the high-disordered sample C demonstrates much lower values of the Zeeman gap.

(ii) The gap values coincide for samples with different gate-gap widths (A1 and A2). This indicates that the equilibration length exceeds our highest gate-gap width ($l_{int}=10\mu$m for the sample A2), justifying the gap determination procedure.

(iii) The gap values within the $\nu_c=1$ incompressible strip significantly exceed the single-particle Zeeman gap in GaAs, demonstrating the exchange enhancement of the gap~\cite{ando}, see Fig.~\ref{gap} (a). The normal field dependence is linear. The slope can be characterized by an effective $g$-factor $g^*\approx7$, however the experimental points are shifted below the straight proportionality.

(iv) There is a pronounced in-plane field dependence of the Zeeman gap,  see  Fig.~\ref{gap} (b). The strength of the dependence is diminishing in higher fields.

Let us start the discussion from   understanding the nature of the measured gap. We detect the flat-band condition $eV=eV_{th}=\Delta_c$ by strong increase of current {\it across} the sample edge. This current is affected by disorder, in contrast to the dissipativeless current along the edges. Energy levels are broadened by  disorder, which is not shown in the simplified Fig.~\ref{sample}. Delocalized states occupy some energy interval around the center of the Landau sublevel. For the strong increase of current it is only necessary to intersect the mobility edges in the two nearby compressible strips. The measured gap  $eV_{th}=\Delta_c$ is  therefore the mobility gap within the incompressible strip, similarly to activation measurements in the bulk~\cite{nichols}. The measured gap should also be characterized by $k=\infty$ since it is detected by charge transport~\cite{ando,nichols}.

This conclusion is confirmed by measurements on samples of different mobilities. The results obtained for high-quality samples A1, A2, and B coincide. In contrast, the high disordered sample C demonstrates much lower gap values, despite the normal magnetic fields equal at least twice  to ones for samples A and B, see Fig~\ref{gap}. The latter indicates higher levels' broadening for high-disordered sample C.  For a high-quality wafer, the mobility is mostly restricted by the long-range potential fluctuations. They have no effect on our local measurements: long-range potential fluctuations can only change the space position of the incompressible strip, which is not seen in the present experiment. Coincidence of the results for the samples of different quality A1, A2, and B indicates saturation of the level broadening for high-quality wafers.  We can conclude that the obtained results are independent on the residual disorder for high-quality samples.

Our main experimental result is the enhancement of the spin gap within $\nu_c=1$ incompressible strip in comparison with the single-particle values, see Fig~\ref{gap} (a). High values of the obtained gap indicate importance of the exchange effects~\cite{ando}. It is well known, that the exchange-enhanced bulk Zeeman splitting  at $\nu=1$ can be characterized~\cite{nichols,aristov,kulik} by high effective $g$-factor. The exact $g^*$-value depends on the experimental method: activation measurements~\cite{nichols}give the highest $g^*=7.3$, while lower $g^*=5$ is obtained from the magnetocapacitance~\cite{aristov}. In the present experiment we study the mobility gap, so the obtained dependence should be compared with the bulk activation measurements~\cite{nichols}, see Fig~\ref{gap} (a). The slopes of both dependencies are close, so the mobility gap behaves similarly in the bulk and at the sample edge. The gap values are  clearly shifted below the direct proportionality, which should be attributed to finite levels broadening~\cite{nichols}. Extrapolation to zero field allows to estimate it roughly as 0.9~meV, indicating higher disorder at the sample edge in comparison with the bulk~\cite{nichols}.

The measured gap can also be characterized by the strong in-plane magnetic field dependence, see  Fig.~\ref{gap} (b).  The in-plane field dependence is usually  attributed~\cite{shmeller,khrapai} to one-particle Zeeman splitting $sg\mu_BB$,  since the exchange term is determined by the perpendicular magnetic field only~\cite{ando}. It allows to obtain~\cite{shmeller} a number of flipped spins $s$, if the relevant single-particle $g$-factor (0.44 or 0.7, see discussion in Refs~\onlinecite{shmeller,khrapai}) is known.
In the present experiment, $sg$ exceeds 0.7 at low $B/B_{\perp}$, see Fig.~\ref{gap} (b).  This should indicate the spin texture (skyrmion) formation with $s>1$ at the sample edge. The obtained number of flipped spins $s$ is diminishing at higher in-plane fields, which also supports the skyrmion picture~\cite{shmeller}. This conclusion is however should be treated with care. The energy level broadening also depends on the in-plane magnetic field~\cite{nichols}, affecting the mobility gap value. These two origins of the strong in-plane field dependence can not be explicitly separated in our experiment.

In summary, we experimentally study the energy gap within the incompressible strip at local filling factor $\nu_c=1$ at the quantum Hall   edge for samples of very different mobilities. The obtained results indicate strong enhancement of the energy gap in comparison to the single-particle Zeeman splitting. We identify the measured gap as a mobility gap, so a pronounced experimental in-plane magnetic  field dependence can both be attributed to the spin effects as well as to the change in the energy levels broadening.

We wish to thank  V.T.~Dolgopolov for fruitful discussions. We gratefully
acknowledge financial support by the RFBR, RAS, the Programme "The
State Support of Leading Scientific Schools".

\end{document}